\documentclass[prl,aps,twocolumn,superscriptaddress,showpacs,floatfix]{revtex4}
\usepackage{amsmath}
\usepackage{amsbsy}
\usepackage{amssymb}
\usepackage[dvips]{graphicx}
\usepackage{epsfig}
\usepackage{color}

\usepackage{psfrag}

\begin{document}

\title{Breathing mode for systems of interacting particles.}

\author{Alain Olivetti}
\email{olivetti@unice.fr}
\author{Julien Barr\'e}
\author{Bruno Marcos}
\affiliation{Laboratoire J. A. Dieudonn\'e, UMR CNRS 6621,
Universit\'e de Nice-Sophia Antipolis, Parc Valrose, F-06108 Nice Cedex 02, France.}

\author{Freddy Bouchet}
\author{Robin Kaiser}
\affiliation{Institut Non-Lin\'eaire de Nice,UMR CNRS 6618,
Universit\'e de Nice-Sophia Antipolis, France. }

\date{\today}

\begin{abstract}

  We study the breathing mode in systems of trapped interacting
  particles. Our approach, based on a dynamical ansatz in the first
  equation of the Bogolyubov-Born-Green-Kirkwood-Yvon (BBGKY)
  hierarchy allows us to tackle at once a wide range of power law
  interactions and interaction strengths, at linear and non linear
  levels. This both puts in a common framework various results
  scattered in the literature, and by widely generalizing these,
  emphasizes universal characters of this breathing mode. Our findings
  are supported by direct numerical simulations.

\end{abstract}

\pacs{05.20.Jj;45.50.-j}

\maketitle

Systems of trapped interacting particles are studied in many areas of
physics: confined plasmas, trapped cold atoms, Bose-Einstein
condensates, colloidal particles, trapped ions, astrophysical systems,
the latter ones being self confined by the interactions.  The
low-lying oscillatory modes of these systems are a natural object of
study, as they are an important non destructive tool to characterize
the system and gain insight into the collective effects at work. As a
consequence, there is an abundant literature on the subject, from the
different areas of research listed above, and corresponding to very
diverse physical situations: (\emph{i}) systems with short range
interactions, such as classical gases or shielded Coulomb interaction
and (\emph{ii}) systems with long range interactions, such as non
neutral plasmas, Coulomb crystals or astrophysical system, in which
the interactions may be weak (gases) or strong (liquids or crystals).

Diverse approaches and techniques are naturally used to investigate
these phenomena.  A trapped classical gas of interacting particles is
studied using a Boltzmann-Vlasov equation in \cite{Guery-Odelin_2002},
where the non linear dynamics is approximated with a scaling ansatz,
which captures the collective effects. Such an ansatz was used earlier
for the Gross-Pitaevskii equation in~\cite{Castin_1996,Kagan_1997}. In
the confined plasma context, the problem is often studied through
hydrodynamical equations, in the so-called ``cold fluid
approximation'' \cite{Dubin_1991}, where the dispersion relation for
fluid modes in a cold spheroidal plasma is derived.  Following an idea
of \cite{Dubin_1993}, Ref.~\cite{Amiranashvili_2003} gives an
approximate solution to the breathing mode of an 1d confined plasma
beyond the cold fluid approximation, using an \emph{ad hoc} closure of
the hydrodynamical equations. Monopole modes of dusty plasmas
interacting with a Yukawa potential are investigated in
\cite{Sheridan_2004_july,Sheridan_2004_december}.  The breathing mode
of trapped ions or colloids interacting \emph{via} Coulomb
interactions has been studied in 1d systems in
\cite{James_1998,Tatarkova_2002} and in 2d in \cite{Partoens_1997} for
crystallized systems, by a direct diagonalization of the linearized
Newtonian equations of motion. Finally, breathing oscillations with
attractive interactions have been studied in an astrophysical context
using the Virial theorem \cite{Chandrasekhar}.

\begin{figure}[th]
\includegraphics[scale=0.4]{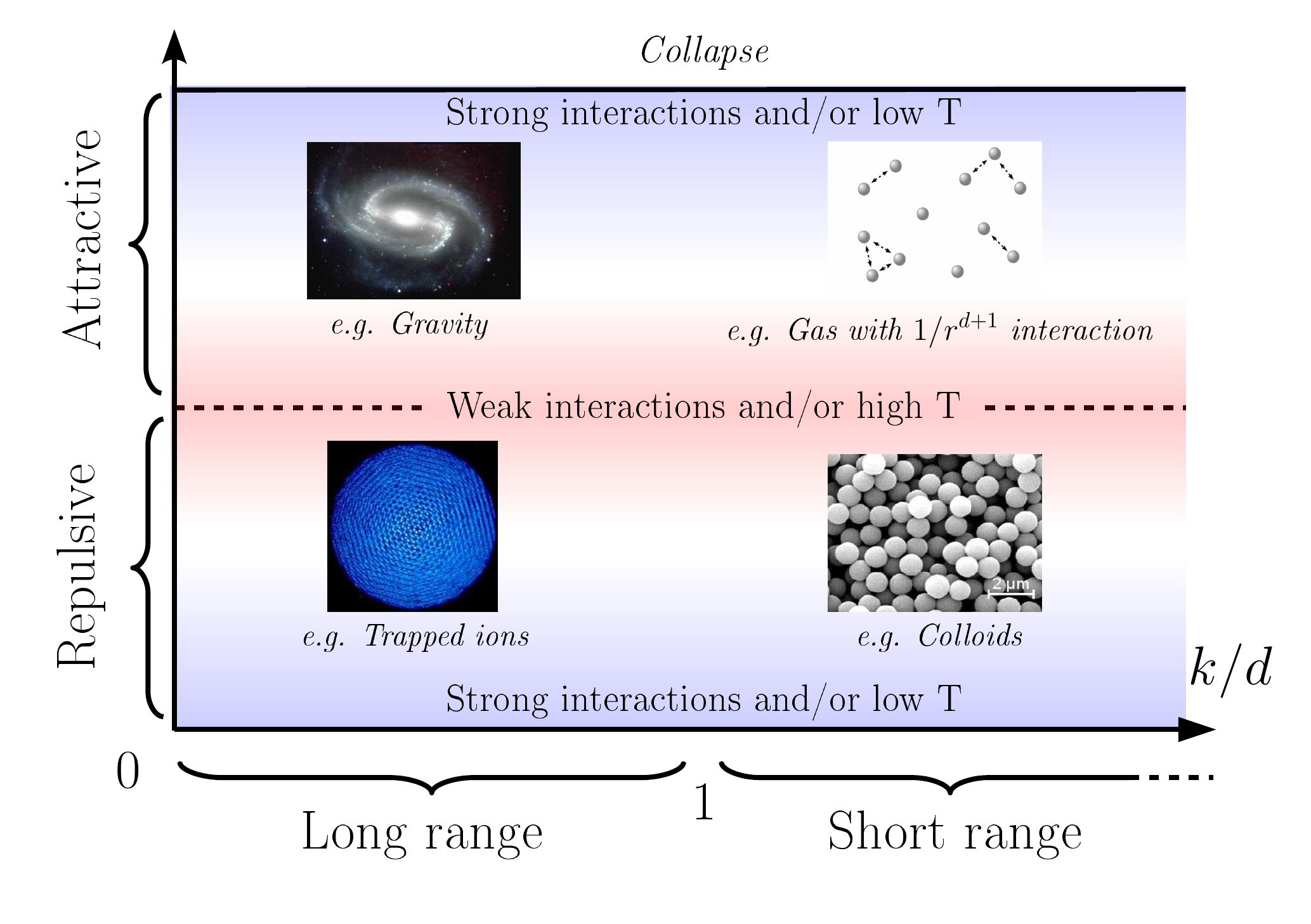}
\caption{(color online). Diagram of the different regimes for the
  breathing mode. On the horizontal axis, the interaction range,
  measured by $k/d$, where $d$ is the dimension of the system. The
  interaction strength is changing along the vertical axis. Pictures
  of some physical examples are inserted for illustration.
  \label{fig1}}
\end{figure}

Each method applies to a specific situation: Newton equations are
adapted to a crystallized state with negligible thermal fluctuations,
linearization assumes a small amplitude, the Vlasov equation is
limited to long range interactions and weak correlations.  Yet in all
cases a similar equation for the breathing mode is obtained. In
particular, it is intriguing that kinetic descriptions assuming small
correlations between particles, fluid descriptions, and perturbative
expansions around a crystallized state all yield similar predictions
for the breathing mode, at linear and non linear levels. This stunning
situation calls for a unified theory. In the limit of zero
temperature, or equivalently infinitely strong interactions, such an
endeavor has recently been undertaken in the linear
regime~\cite{Henning_2008}.  A more general situation summarizing the
different possible regimes for a binary isotropic power-law
interparticle force $F(r)\sim 1/r^k$ is shown in a diagram
Fig.~\ref{fig1}. We have organized the different cases along two axis.
On the horizontal axis we represent the interaction range, which we
will call long-range if $k/d\leq 1$ and short range otherwise. The
case $k/d \leq 1$ corresponds to non integrable forces at large
distances \cite{ShortLongRange}.  The vertical axis represents the
interaction strength with respect to the thermal energy.

In this work, we present a theory of the breathing mode of systems of
classical trapped interacting particles which classifies many cases
studied in the references cited above in a common framework. The
theory is valid both for short-range and long-range interactions, for
any dimension, and for various interaction strengths: it applies to
the whole diagram of Fig.~\ref{fig1}, with restrictions only for
strongly attractive long range interactions and for even moderate
attractive short range interactions, where gravitational like
collapses in the former case, and strong instabilities due to the
unregularized short range singularity in the latter are expected.  Our
theory describes both linear as well as non linear oscillations, and
isolated systems as well as systems in contact with a thermal bath.

We consider a system of particles confined by an harmonic spherical
trapping force $\textbf{F}_{trap}(\textbf{r})=-\omega^2_0\textbf{r}$,
with binary interaction forces $\textbf{F}_{int}$. In the
canonical setting, particles are subjected to a positive constant friction
$\kappa$ and diffusion $D$.  In the microcanonical setting,
$\kappa=0$, $D=0$ and the dynamics is Hamiltonian. To overcome the
limitations in the validity of the Vlasov equation, we describe the
cloud of particles by its one-particle and two-particles distribution
functions $f(\textbf{r}_{1},\textbf{v}_{1},t)$ and
$g(\textbf{r}_1,\textbf{v}_1,\textbf{r}_2,\textbf{v}_2,t)$. We
start from the first equation of the
BBGKY hierarchy, which we complement  by a Fokker-Planck
operator to include the
temperature in the canonical case:
\begin{equation}\label{Equation_BBGKY+FP}
\frac{\partial f}{\partial t}+\nabla_{\textbf{r}}.(\textbf{v}f)
+\textbf{F}_{trap}. \nabla_{\textbf{v}}f+C[g]=
D\Delta_{\textbf{v}}f+\kappa\nabla_{\textbf{v}}.(\textbf{v}f),
\end{equation}
where $C[g]$ is the interaction term given by:
\begin{equation}\label{Collision_operator}
  C[g](\textbf{r}_1,\textbf{v}_1,t)=\int \textbf{F}_{int}(\textbf{r}_1,\textbf{r}).\nabla_{\textbf{v}_1}g(\textbf{r}_1,\textbf{v}_1,\textbf{r},\textbf{v},t)\,d\textbf{r}d\textbf{v}.
\end{equation}
We stress that Eq.~(\ref{Equation_BBGKY+FP}), in contrast with the
Vlasov equation, can also describe strongly correlated systems.  We
assume in the following the existence of a stationary state $f_0$ and
$g_0$, not necessarily the thermodynamic equilibrium~\cite{Houches}.
We now drastically simplify the dynamics by using a scaling ansatz
\cite{Castin_1996,Kagan_1997,Guery-Odelin_2002}, which we extend here
to the two-particles function $g$:
\begin{equation}\label{Equation_Ansatz}
  \left\{\begin{array}{l}
      f(\textbf{r}_1,\textbf{v}_1,t)=f_0(\varphi(\textbf{r}_1,\textbf{v}_1))\\
      g(\textbf{r}_1,\textbf{v}_1,\textbf{r}_2,\textbf{v}_2,t)=g_0(\psi(\textbf{r}_1,\textbf{v}_1,\textbf{r}_2,\textbf{v}_2))
    \end{array}\right.
\end{equation}
with
$\varphi(\textbf{r}_1,\textbf{v}_1)=(\textbf{R}_1=\textbf{r}_1/\lambda,
\textbf{V}_1=\lambda \textbf{v}_1-\dot{\lambda} \textbf{r}_1)$ and
$\psi(\textbf{r}_1,\textbf{v}_1,\textbf{r}_2,\textbf{v}_2)=(\varphi(\textbf{r}_1,\textbf{v}_1),\varphi(\textbf{r}_2,\textbf{v}_2))$.
All time dependence in the dynamics is now included in the positive
parameter $\lambda$. Introducing Eq.~(\ref{Equation_Ansatz}) into
Eq.~(\ref{Equation_BBGKY+FP}) leads to:
\begin{equation}\label{Equation_intermediaire_1}
\begin{array}{l}
  \underset{i=1}{\overset{d}{\sum}} \left\{ \frac{V_i}{\lambda^2}\frac{\partial f_0}{\partial R_i}-R_i\lambda\frac{\partial f_0}{\partial V_i}(\ddot{\lambda}+\omega_0^2\lambda)
    -\kappa\frac{\partial V_if_0}{\partial V_i}-\kappa\lambda\dot{\lambda}R_i\frac{\partial f_0}{\partial V_i}\right.\\
  \left.-D\lambda^{2}\frac{\partial^2 f_0}{\partial V_i^2}\right\}+
  C[g_0\circ \psi](\textbf{r}_1,\textbf{v}_1,t)=0,
\end{array}
\end{equation}
where the difficulty is to deal with the interaction term. We now assume
a homogeneous two-body interaction with degree $-k$:
\begin{equation}\label{Hyp_homogeneite}
  \textbf{F}_{int}(\lambda\textbf{r}_1,\lambda\textbf{r}_2)=\frac{1}{\lambda^k}
\textbf{F}_{int}(\textbf{r}_1,\textbf{r}_2),
\end{equation}
as for example a pure power law.
The important step is to replace the interaction term
$C[g_0\circ\psi](\textbf{r}_1,\textbf{v}_1,t)$ by a linear combination
of $f_0$ and its derivatives. This is achieved using the condition
\eqref{Hyp_homogeneite} and the fact that $f_0$ and $g_0$ are
stationary solutions of Eq.~(\ref{Equation_BBGKY+FP}).
Equation~\eqref{Equation_intermediaire_1} becomes
\begin{equation}\label{Equation_intermediaire_2}
\begin{array}{l}
  \underset{i=1}{\overset{d}{\sum}}\left\{V_i\frac{\partial f_0}{\partial R_i}\left(\frac{1}{\lambda^2}-\lambda^{1-k}\right)+D\frac{\partial^2 f_0}{\partial V_i^2}(\lambda^{1-k}-\lambda^2)\right.\\
  -R_i\frac{\partial f_0}{\partial V_i}\left[\lambda\left(\ddot{\lambda}+\omega_0^2\lambda\right)-\lambda^{1-k}\omega_0^2+\kappa\lambda\dot{\lambda}\right]\\
  \left.+\kappa\frac{\partial V_if_0}{\partial V_i}(\lambda^{1-k}-1)\right\}=0.
\end{array}
\end{equation}
Multiplying the previous equation by $R_jV_j/N$, and integrating over
$d\textbf{R}d\textbf{V}$, we obtain a constraint on the parameter
$\lambda$:
\begin{equation}\label{Equation_oscillation_1}
  \ddot{\lambda}+\kappa\dot{\lambda}+\left(\lambda-\frac{1}{\lambda^{k}}\right)\omega_0^2-\left(\frac{1}{\lambda^3}-\frac{1}{\lambda^{k}}\right)\frac{\langle V_j^2\rangle_{f_0}}{\langle R_j^2\rangle_{f_0}}=0,
\end{equation}
where $j$ is a coordinate label, and we have set $ \langle \chi
\rangle_f=\frac{1}{N}\int \chi(\textbf{r},\textbf{v})
f(\textbf{r},\textbf{v},t)d\textbf{r}d\textbf{v} $.  In the dynamical
equation for $\lambda$ (Eq.~(\ref{Equation_oscillation_1})), all
parameters are computed as averages over the stationary distribution
$f_0$. For Eq. (\ref{Equation_oscillation_1}) to be a unique equation,
it is necessary that the ratio $\langle V_j^2\rangle_{f_0}/\langle
R_j^2\rangle_{f_0}$ does not depend on~$j$, which is true if the trap
and interactions are isotropic.

We introduce the dimensionless parameter $p= \langle
V_j^2\rangle_{f_0}/(\omega_0^2\langle R_j^2\rangle_{f_0})\sim
k_{B}T/E_{trap}$, wher $k_{B}T$ is the thermal energy and $E_{trap}$ the
typical potential energy due to the trap. At the canonical
equilibrium, $\langle V_j^2\rangle_{f_0}=\omega_0^2 L^2$, where $L$ is
the typical size of the system without interaction. The parameter
$p=L^2/\langle R_j^2\rangle_{f_0}$ thus describes change of the
square of the size of the trap due to the interactions.
The range $p<1$ (resp.  $p>1$) corresponds
to a repulsive (resp. attractive) interaction. A value of the
parameter $p\sim 1$ means high temperature or negligible
interactions. The limits $p\to 0$ and $p\to +\infty$ correspond to zero
temperature or strong repulsive and attractive interaction.
We can now rewrite Eq.~(\ref{Equation_oscillation_1}) as

\begin{equation}
\ddot{\lambda}+\kappa\dot{\lambda}+\phi'(\lambda)=0,
\end{equation}
which corresponds to the
equation of a damped anharmonic oscillator in the potential $\phi$:
\begin{equation}
\label{phi}
\phi(\lambda)=\left\{
\begin{array}{ll}
\omega_0^2\left(\frac{1}{2}\lambda^2+\frac{1}{2}\frac{p}{\lambda^2}+\frac{p-1}{1-k}\lambda^{1-k}\right), & \text{if }k\neq 1,\\
\omega_0^2\left(\frac{1}{2}\lambda^2+\frac{1}{2}\frac{p}{\lambda^2}+(p-1)\log \lambda\right),&\text{if } k=1~.
\end{array}
\right.
\end{equation}
The first term in Eq.~\eqref{phi} is the quadratic confining
potential, the second one corresponds to a pressure term and the last
one is introduced by the two-body interaction.

For repulsive interactions ($p<1$), $\phi$ is strictly convex for all
$k\geqslant 0$. It diverges as $\lambda^{-2}$ when $\lambda \to 0$ and
as $\omega_{0}^2\lambda^2/2$ when $\lambda \to +\infty$.  Its unique
minimum is $\lambda=1$. The $\lambda^{-2}$ divergence at small
$\lambda$ is due to pressure effects for very compressed clouds, and
thus does not depend on the interaction. It yields a generic shape for
the breathing oscillations in the nonlinear regime. For attractive
interactions ($p>1$), if $0\leqslant k\leqslant 3$, $\phi$ has exactly
the same qualitative properties as in the repulsive case. For $k>3$,
$\phi$ tends to minus infinity when $\lambda$ goes to zero, indicating
a possible collapse of the cloud. However, such unregularized power
law forces may create instabilities, so that these predictions for the
breathing dynamics should be considered with care.
\begin{figure}[th]
\includegraphics[scale=0.4]{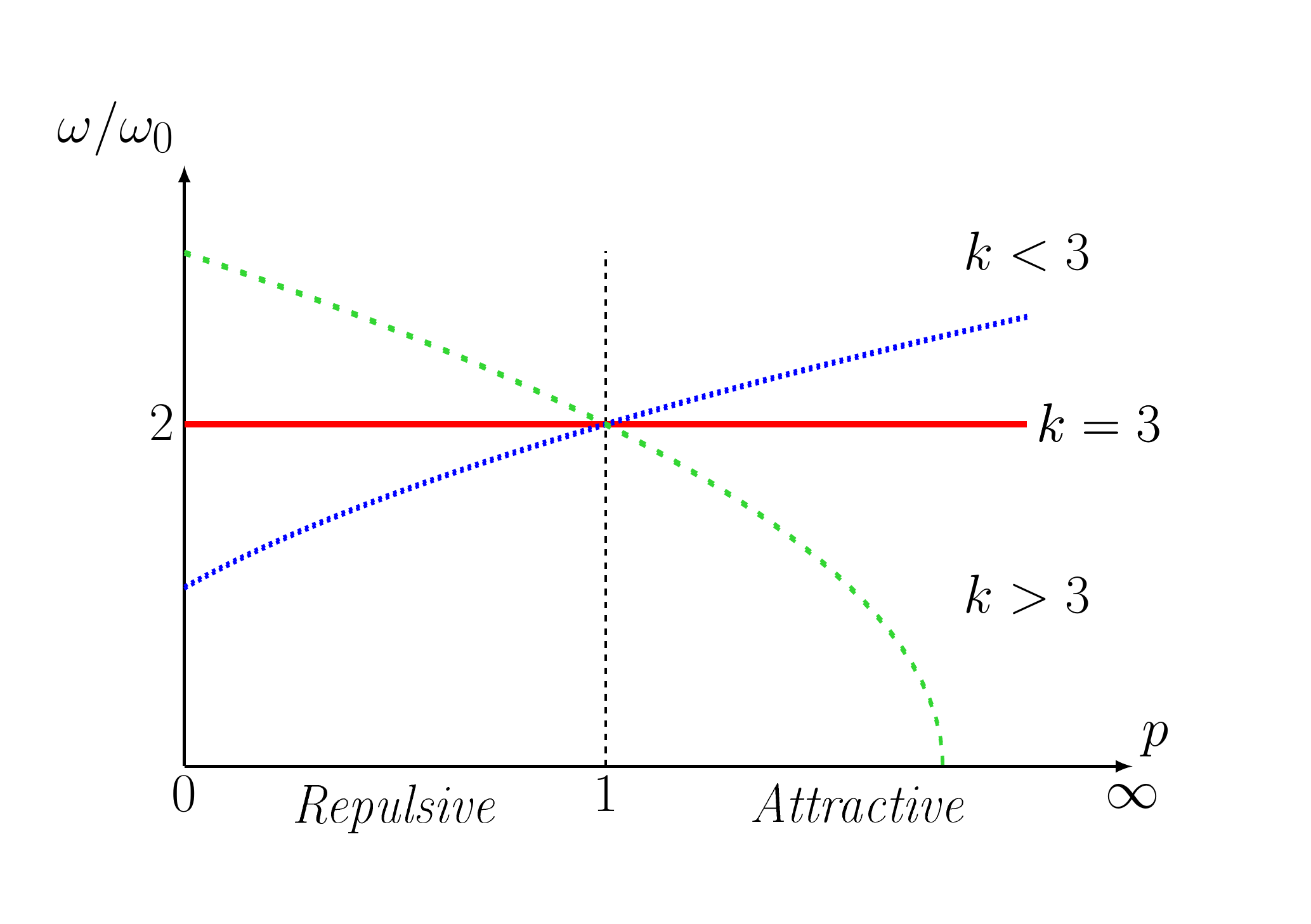}
\caption{(color online). Frequency of the linearized breathing mode as
  a function of the interaction strength $p$, for different values of
  interaction range $k$.\label{fig:omega}}
\end{figure}

From Eq.~\eqref{phi}, we obtain the general expression of the breathing
oscillation frequency in the small friction limit, as a function of
the interaction range $k$ and the interaction strength $p$:
\begin{equation}
\label{eq:freq_lin}
\omega(k,p)=\omega_0\left[(3-k)(p-1)+4 \right]^{1/2}.
\end{equation}
This expression recovers the well known limits $\omega=2\omega_0$ for
a non interacting gas ($p=1$) and $\omega=\sqrt{3}\omega_0$ for a
strongly interacting Coulomb plasma ($p=0$, $k=2$). It provides a
generalization to the whole $(k,p)$ plane shown in
Fig.~\ref{fig:omega} and is independent of the dimension.  We note
that in 3 dimensions, the breathing frequency is a decreasing function
of the interaction strength for repulsive long range interactions, and
a increasing function of the interaction strength for repulsive short
range interaction.

We can now compare the general Eq.~(\ref{Equation_oscillation_1}) to
the results found in the literature for various specific situations.
Oscillations of crystallized systems \cite{James_1998,Tatarkova_2002,
  Partoens_1997,Henning_2008} correspond to negligible pressure
effects, i.e. $p=0$ and the $\lambda^{-3}$ term of is absent. In
\cite{Amiranashvili_2003}, the authors consider a 1d plasma ($k=0$)
with $p$ not too small, and introduce a pressure yielding the
$\lambda^{-3}$ term, which leads to the exact equivalent of
Eq.~(\ref{Equation_oscillation_1}).  Note that
Eq.~(\ref{Equation_oscillation_1}) also contains the case of a
classical gas with "mean field" interactions \cite{Guery-Odelin_2002}.
This work considers a Dirac $\delta$ potential, which corresponds to a
homogeneity degree $-k=-d-1$. This result emphasizes that the present
theory is not only valid for power-law forces.

\begin{figure}[th]
\includegraphics[scale=0.4]{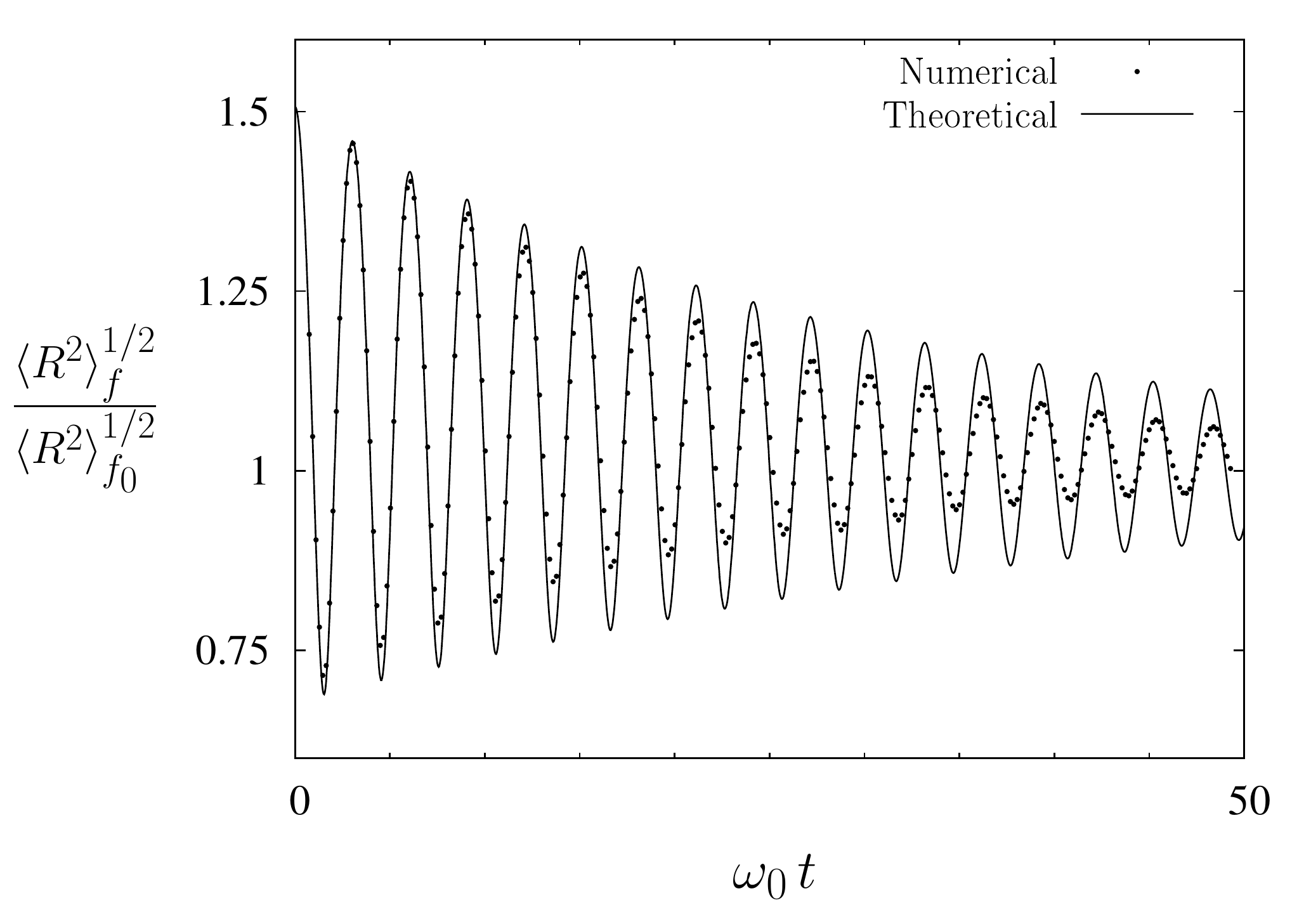}
\caption{Evolution of the typical size of the cloud, in a case where
  the ansatz does not describe the full dynamics. The space dimension
  is $d=2$, and the interactions are repulsive. The parameters are
  $k=4$ (short range interaction), $\omega_0/\kappa=17.8$ and
  $p=0.63$. \label{fig3}}
\end{figure}

\begin{figure}[th]
\includegraphics[scale=0.4]{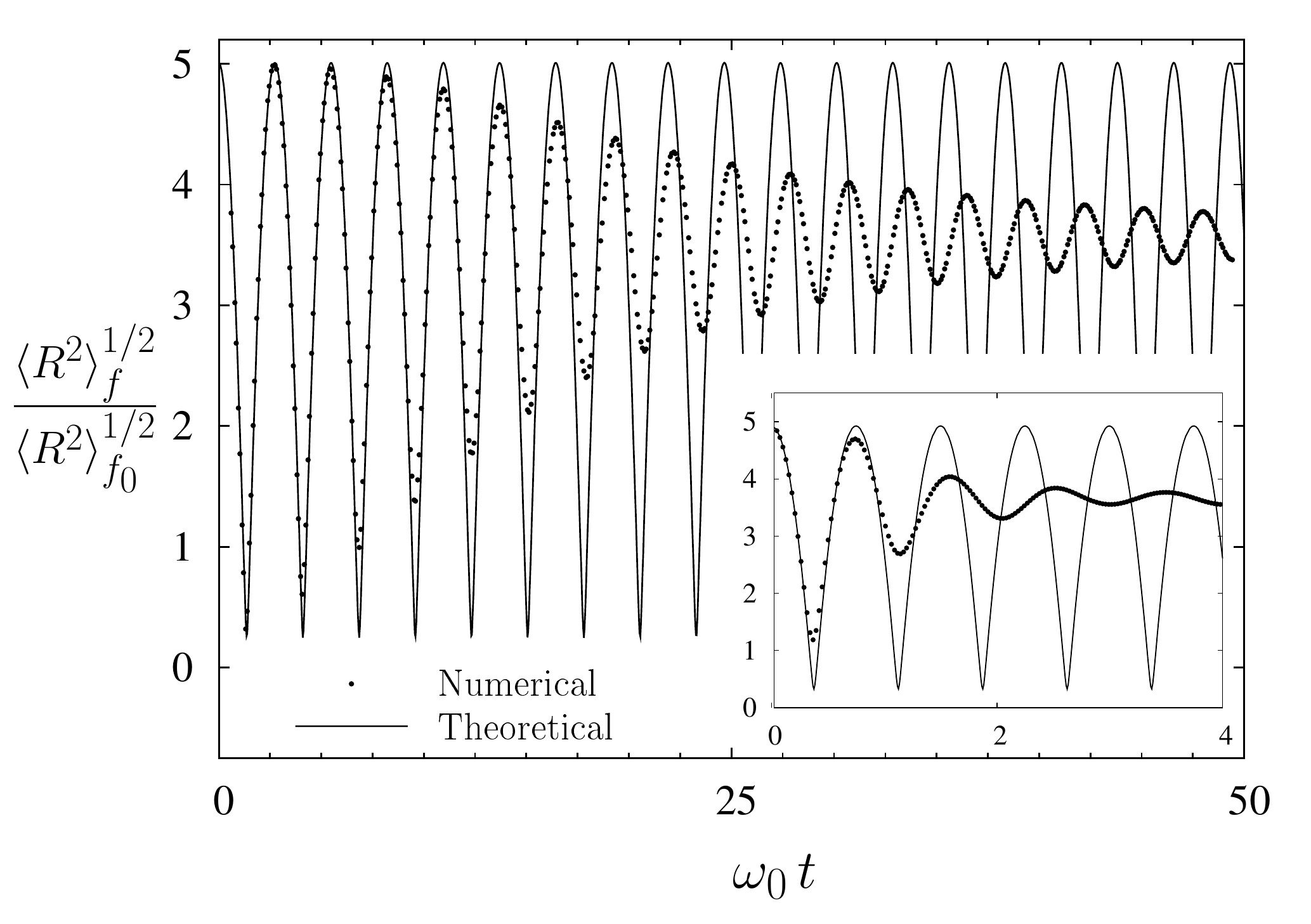}
\caption{Evolution of the typical size of the cloud, in a case where
  the ansatz does not describe the full dynamics. The frequency is
  still very well captured. The space dimension is $d=3$, and the
  interactions are attractive. The parameters are $k=0$ (long range
  interaction), $\omega_0=17.8$, $\kappa=0$ (microcanonical ensemble)
  and $p=2.2$. Same parameters for the inset except $p=70$.  \label{fig4}}
\end{figure}

In order to test the domain of validity of the ansatz solution, we
have performed numerical simulations for different force index $k$,
parameter $p$ and amplitude of initial perturbation, in two and three
dimensions, with (canonical ensemble) or without (microcanonical
ensemble) a thermostat.  We simulate the system using a molecular
dynamics approach with $N=4000$ particles.  The integrator scheme is a
Verlet-leapfrog algorithm \cite{Allen_1987} in the microcanonical or
canonical ensemble. The forces are exactly computed at each time-step.
As strong short range singularities for parameters in the upper right
corner of Fig.~\ref{fig1} create numerical difficulties, we have not
tested the theory in this region. The computer simulations are
performed as follows: we first equilibrate the system in a stationary
state $f_0$.  Then, at $t=0$, we introduce a perturbation by rescaling
the positions and velocities according to Eq.~(\ref{Equation_Ansatz})
and we let the system evolve.  A similar simulation of a 1d Coulomb
system in the microcanonical ensemble has been performed in
\cite{Amiranashvili_2003}. The results of our extensive simulations
may be summarized as follows: (\emph{i})
Eq.~(\ref{Equation_oscillation_1}) always picks up quite precisely the
oscillation frequency, but not always the amplitude decay.
(\emph{ii}) For strongly repulsive interaction ($p\to0$),
Eq.~(\ref{Equation_oscillation_1}) describes very precisely the whole
dynamics. (\emph{iii}) For a repulsive long-range or short range
interaction and intermediate $p$ (i.e. $p\sim 0.5$) the agreement for
the oscillation amplitude is not perfect (Fig.~\ref{fig3}).
(\emph{iv}) For attractive long range interactions, the accuracy of
the ansatz degrades as $p$ increases (Fig.~\ref{fig4}).

To explain these results, we first stress that in the limit $p\to0$,
Eq.~(\ref{Equation_oscillation_1}) is {\em exact}. In this case, it
may indeed be derived directly from Newton equations, as done in
\cite{Henning_2008} in the linear approximation.  The correct
generalization for an arbitrary perturbation amplitude is given by
Eq.~(\ref{Equation_oscillation_1}). For intermediate $p$, we attribute
the discrepancy between the predicted and simulated oscillation
amplitudes to effects that are not taken into account in the simple
dynamical ansatz~(\ref{Equation_Ansatz}), and thus limit the validity
of Eq.~(\ref{Equation_oscillation_1}).  Indeed, for long range
interactions, one would expect collective effects neglected in the
ansatz (Landau damping, phase mixing, etc) to play a role in the
oscillation decay beyond the friction $\kappa$). Similarly, for short
range interactions, two-body collisions may be important.  This
explanation is supported by the frictionless microcanonical
simulations: when there is no amplitude decay in the microcanonical
ensemble, which means that phase mixing and two body collisions are
negligible, Eq.~(\ref{Equation_oscillation_1}) correctly predicts the
breathing frequency and amplitude, with or without friction.
Conversely, amplitude decay or modulation in the microcanonical
ensemble is associated with discrepancies between theory and
simulations.

In summary, starting from the first equation of the BBGKY hierarchy
and a scaling ansatz for the dynamics, we have derived a non-linear
equation describing the breathing oscillations of trapped particles
interacting \emph{via} homogeneous forces. The derivation and equation
are valid independently of the temperature, interaction strength,
interaction range and dimensionality of the physical space, and it is
successfully confronted to direct numerical simulations. The main
limitation is due to phase mixing phenomena for long range interacting
systems and two-body collisions in short range interacting ones, especially
for weak repulsive and attractive interactions, where they introduce
damping and loss of coherence, unaccounted for in the scaling ansatz.
Even though we have concentrated on power-law interactions, the
homogeneity condition for the force is more general. It includes for
instance dipolar interactions and some non potential forces, allowing
the use of the ansatz technique in such cases. Finally, the general
results obtained in this letter might be a useful guide for
experimentalists to extract information on the interaction between
particles from easily measurable phenomena, such as breathing mode
dynamics. 

Beyond the breathing mode, a generic study of quadrupolar modes would
be very desirable, as harmonic traps are often anisotropic in
experimental situations. This is not possible with the scaling ansatz,
except in special cases. However, following the lines of the present
letter, and applying methods used in~\cite{Sheridan_2004_december}, a
more general approach should be possible.


\begin{thebibliography}{99}
\bibitem{ShortLongRange} The
  boundary between long and short range for equilibrium statistical
  properties is $(k-1)/d=1$, which corresponds to the integrability
  limit of the interaction potential~\cite{Houches}. For dynamical
  properties, it seems preferable to use the present definition.
\end{thebibliography}
\end{document}